\documentclass[manuscript]{emulateapj}

\slugcomment{to appear in the Astrophysical Journal}

\shorttitle{Recurrent Nova Model}
\shortauthors{Kato et al.}

\usepackage{apjfonts}

\begin{document}


\title{A Millennium-Long Evolution of One-Year-Recurrence-Period Nova
-- Search for Any Indication of the Forthcoming H\lowercase{e} Flash
}

\author{Mariko Kato} 
\affil{Department of Astronomy, Keio University, Hiyoshi, Yokohama
  223-8521, Japan;}
\email{mariko.kato@hc.st.keio.ac.jp}

\author{Hideyuki Saio}
\affil{Astronomical Institute, Graduate School of Science,
    Tohoku University, Sendai, 980-8578, Japan}

\and
\author{Izumi Hachisu}
\affil{Department of Earth Science and Astronomy, College of Arts and
Sciences, The University of Tokyo, 3-8-1 Komaba, Meguro-ku,
Tokyo 153-8902, Japan}

\begin{abstract} 
We present 1500 cycles of hydrogen shell flashes on a $1.38~M_\sun$ white 
dwarf (WD) for a mass accretion rate of 
$1.6 \times 10^{-7}~M_\sun$~yr$^{-1}$, the mass ejection
of which is calculated consistently with the optically thick winds.
This model mimics the one-year-recurrence-period nova M31N 2008-12a. 
Through these hydrogen flashes a helium ash layer grows in mass 
and eventually triggers a helium nova outburst.  
Each hydrogen flash is almost identical and there is no precursor
for the forthcoming He flash either in the outburst or in the quiescent 
until the next He flash suddenly occurs. 
Thus, M31N 2008-12a is a promising candidate of He novae,
outbursting in any time within a millennium years. 
The prompt X-ray flash of He nova lasts as short as 15 min
with the X-ray luminosity being about a half of the Eddington luminosity, 
making the observation difficult. 
In the very early phase of a He flash, the uppermost H-rich layer is 
convectively mixed into the deep interior and most of hydrogen is consumed 
by nuclear burning.  
In comparison with hydrogen shell flashes of M31N 2008-12a,
we expect the forthcoming He nova with a very short prompt
X-ray flash (15 min), a very bright optical/NIR peak 
($\sim3.5$ mag brighter than M31N 2008-12a), a much longer nova duration 
($>2$ years), and a longer supersoft X-ray source 
phase (40-50 days or more).
\end{abstract}


\section{Introduction}
\label{sec_introduction}

A nova is a hydrogen flash on a mass-accreting white dwarf (WD)
\citep{nar80,ibe82,pri86,sio79,spa78}.
Multicycle nova outbursts have been calculated
with Henyey-type evolution codes.  Those codes, however,
meet numerical difficulties when the nova envelope expands to a giant size.
To continue the numerical calculation beyond this stage,
various authors have adopted various mass-loss schemes and approximations
\citep{pri95,kov98,den13,ida13,wol13a,wol13b,kat14shn,kat15sh,tan14}.
In the previous paper \citep{kat17sh}, we established 
an iteration method for calculating the extended stage of novae with 
time-dependent mass-loss rates of optically thick winds, and  
presented a model for one full cycle of a nova outburst 
for the recurrent nova M31N 2008-12a. 

M31N 2008-12a has exploded almost every year which makes this 
object as the shortest record of the recurrence period of 
$P_{\rm rec}\sim 1$~yr
\citep{dar14,dar15,dar16errata, hen14, hen15,tan14,dar16} or 
$P_{\rm rec}\sim 0.5$ yr.\footnote{\citet{hen15.half.period} proposed 
   a 0.5 yr recurrence period to explain the discrepancy between
   the early (around 2000) and recent trends of the outburst cycles.
   Even if an outburst occurred in the middle of the 1 yr cycle,
   we could not observe it due to Sun constraint.} 
\citet{kat15sh,kat17sh} presented the outburst model 
of a $1.38~M_\sun$ WD with a mass-accretion rate of 
$\dot M_{\rm acc} = 1.6 \times 10^{-7}~M_\odot$~yr$^{-1}$. 
During the outburst of the model, a part of the envelope mass
is blown in the optically thick wind, and the rest is processed 
to helium and accumulates on the WD. 
Thus, we expect that, after many flashes, 
the helium mass gradually increases and eventually reaches a critical value
for ignition, leading to a He flash. 
In other words, M31N 2008-12a is a promising candidate of He novae.

Helium novae were theoretically predicted by \citet{kat89} as a
nova explosion caused by a helium shell flash on a WD. 
Binary systems of He nova progenitors were categorized into three types 
\citep{kat89,kat08}.\\
(1) WDs accreting helium matter from a helium star companion. \\
(2) WDs accreting hydrogen with rates high enough to keep steady 
hydrogen burning, i.e., the accretion rate is higher than the stability line 
\citep{kat14shn}. 
Such objects correspond to persistent supersoft X-ray sources.\\ 
(3) WDs accreting hydrogen with rates lower than the stability line, 
but high enough to increase the helium layer mass. 
Such objects correspond to recurrent novae. 

Kato et al.'s (1989) prediction was realized as a type (1) when
V445 Pup was discovered on UT 2000 December 30 by K. Kanatsu \citep{kan00}. 
V445 Pup is the first and only identified helium nova that underwent 
a helium shell flash \citep{kam02,kat03,ash03,iij08,kat08,wou09}.
The other two types of helium nova systems were not detected yet. 
As mentioned above, the recurrent nova M31N 2008-12a 
is a promising candidate of type (3) He novae.  
As M31N 2008-12a is the shortest recurrence period nova, 
we expect that the He layer is now growing in mass at high rates that 
result in He ignition in the near future.  
Thus, the theoretical description for He nova outbursts could be useful
for making observational plans. 

A successive shell flash calculation is not easy, because  
we need substantial computer resources. Only 
few groups have ever presented such calculations.
\citet{epe07} calculated $>1000$ successive cycles of nova outbursts 
on a $0.65~M_\sun$ WD for the mass-accretion rate of 
$\dot M_{\rm acc}= 1\times 10^{-9}M_\sun$~yr$^{-1}$ and on a 
$1.0~M_\sun$ WD for $\dot M_{\rm acc}= 1\times 10^{-11}M_\sun$~yr$^{-1}$. 
In these models, the WDs are eroded in 
every outburst and the WD masses secularly decrease. 
As the He ash is lost, no He flashes occur. 

\citet{ida13} calculated a few thousand successive hydrogen flashes 
for 1.0, 1.25, 1.35, and 1.4 $M_\sun$ WDs with 
$\dot M_{\rm acc}=1 \times 10^{-6}M_\sun$~yr$^{-1}$, 
and showed that the He layers grow in mass to result in He flashes. 
This accretion rate is much higher than that of the stability line, 
i.e., hydrogen should steadily burn without flashes, 
unless the accretion is stopped and restarted artificially.   
We call such novae ``the forced novae'' 
\citep[see ][]{kat14shn,hac16sk}.  
This accretion rate is too high to be applicable to M31N 2008-12a.

\citet{hil16} also presented successive hydrogen flashes 
starting from a hot $1.34~M_\sun$ WD with 
$\dot M_{\rm acc}\sim 1 \times 10^{-7}M_\sun$~yr$^{-1}$, 
which result in a He flash after 2573 cycles of hydrogen flashes.  
The recurrence period of the H flashes is about 2 years, close to 
1 yr of M31N 2008-12a. 
These authors, however, paid little attention to 
describing the flash properties in detail. 
Moreover, the adopted parameters are not appropriate for 
M31N 2008-12a, which makes difficult drawing  
practical information for observation. 

The aim of this work is to calculate a number of hydrogen flashes until 
a He flash occurs for the appropriate parameters of M31N 2008-12a. 
This paper is organized as follows.  
Section \ref{sec_method} introduces our numerical method and 
various parameters for the model. 
Section \ref{sec_Hflash} describes physical properties in 
our thousands of nova outbursts. 
Section \ref{sec_Heflash} shows the physical properties of 
the early phase of helium burning. 
Discussion and conclusions follow in Sections \ref{sec_discussion}
and \ref{sec_conclusion}.


\begin{deluxetable}{llll}
\tabletypesize{\scriptsize}
\tablecaption{Initial Model Parameters} 
\tablewidth{0pt}
\tablehead{
\colhead{Subject}&
\colhead{units}&
\colhead{quantity}
}
\startdata
$M_{\rm WD}$ & $M_\odot$ & 1.380\\
$\log R_{\rm WD}$ & $R_\odot$ & $-2.572$ \\
$\log L_{\rm WD}$ & $L_\odot$ & 2.00 \\
$\log g$ & cm~s$^{-2}$ & 9.72 \\
$\log V_{\rm esc}$ & cm~s$^{-1}$ & 9.15\\
$\log T_{\rm c}\tablenotemark{a}$ & K  & 8.03   \\
$\dot M_{\rm acc}$ & $10^{-7}~M_\odot$~yr$^{-1}$ & 1.6\\
$M_{\rm H~env}$ & $10^{-8}~M_\odot$  &6.47\\
$\log L_{\rm ph}$ & $L_\odot$ & 4.08 \\
$\log L_{\rm nuc}$ & $L_\odot$ & 4.05 \\
$\log T_{\rm ph}$ & K & 6.05 \\
$\log R_{\rm ph}$ & $R_\odot$ & $-2.54$
\enddata
\tablenotetext{a}{The temperature at the WD center.}
\label{table_initial}
\end{deluxetable}

\section{Numerical Method}
\label{sec_method}

We have calculated 1500 consecutive nova outbursts 
leading to a helium flash. 
We adopt the WD mass of $1.38~M_\sun$ and the mass-accretion rate of  
$1.6 \times 10^{-7}~M_\sun$~yr$^{-1}$, taken from the model for the  
one-year-recurrence-period nova M31N 2008-12a \citep{kat16xflash}. 
We use the same Henyey-type code as in the previous work 
\citep{kat16xflash,kat17sh}. 
The chemical compositions of the accreting matter and initial hydrogen-rich
envelope of the WD are assumed to be $X=0.7, ~Y=0.28$, and $Z=0.02$.
To save computer time, we use a small nuclear reaction network 
up to magnesium. 
Convective mixing is treated diffusively adopting the effective diffusion 
coefficient derived by \citet{spr92}. 
 Although the coefficient was derived for semiconvective mixing
 (corresponding to the Schwarzschild type), it leads to a uniform
  chemical composition distribution in the fully convective zone.
 Therefore, we use the coefficient whenever radiative temperature
 gradient exceeds adiabatic temperature gradient 
 \citep[Schwarzschild-Kato criterion,][for the occurrence of 
 convection]{kat66}. 
Neither convective overshooting nor diffusion processes of nuclei are
included. 
We neglect the effects of rotation for simplicity. 
Including rotation into the evolution code is difficult and beyond the 
scope of the present work. 
Also, accretion energy outside the photosphere is not included. 
For technical simplicity, we assume that gas is accreted with the same
temperature as the stellar surface. 
These assumptions are discussed in Section \ref{sec_accenergy}.

It should be noted that the quiescent luminosity, $L_{\rm ph}$, which 
appears in Figures \ref{12year}, \ref{Levol}, 
\ref{Levolnuc}, and \ref{hr}, cannot be directly compared with observation  
because it does not include the accretion energy outside the photosphere
that is reprocessed with the accretion disk surface 
\citep[e.g.,][]{hkkm00, hac01kb, hac06kb}.

During the extended stages of nova outbursts, 
the optically-thick wind mass-loss occurs. 
\citet{kat17sh} presented two series of time-dependent 
wind mass-loss rates  
for a nova outburst model of M31N 2008-12a. 
To avoid time consuming process of iterative numerical fitting with 
optically thick wind solutions in each stage, 
we here simply assumed wind mass-loss rates in our Henyey-code 
calculation and followed a few thousand flashes. 
The adopted mass-loss rates 
\citep[the dotted green line in Figure 11 in][]{kat17sh}
reasonably mimic those of the optically thick winds, 
but are slightly overestimated. 

We adopt $> 5000$ mass zones to cover the entire configuration  
including a carbon-oxygen (CO) WD core, a He layer, and a H-rich envelope. 
Such a large number of (i.e., very fine) mass zones are necessary to guarantee 
numerical accuracy especially in rapidly changing 
physical variables, such as 
the temperature, density, and chemical composition of nuclear burning 
region and also expanding region. 
Rezoning is adopted when it is necessary in a way to conserve 
mass, energy, and chemical composition. 
The time step is chosen to be short enough ($ < 4\times 10^4$ seconds) 
in calculation of the 1500 cycles of hydrogen shell flashes, 
but much shorter (1 second) for the He ignition. 
It took about a week of CPU time on a PC (Xeon E5-1660, 3.70 GHz)
for entire sequence of 1500 hydrogen shell flashes followed by the He flash
until we stop calculation. 

We adopted an initial WD model 
in which an energy balance is already established 
between heating (by the mass accretion and nuclear energy generation)  
and cooling (by the radiative transfer and neutrino
energy loss) \citep{kat14shn}. 
This is a good approximation of the long time-averaged evolution 
of a mass accreting WD. Starting from 
such an equilibrium state, the nova cycle approaches a limit cycle 
\citep{kat14shn} in a short time. 
We will discuss in more detail on the initial model 
in Section \ref{sec_initialmodel}.  
Parameters of our initial WD model is summarized in Table \ref{table_initial}.


\begin{figure}
\epsscale{1.1}
\plotone{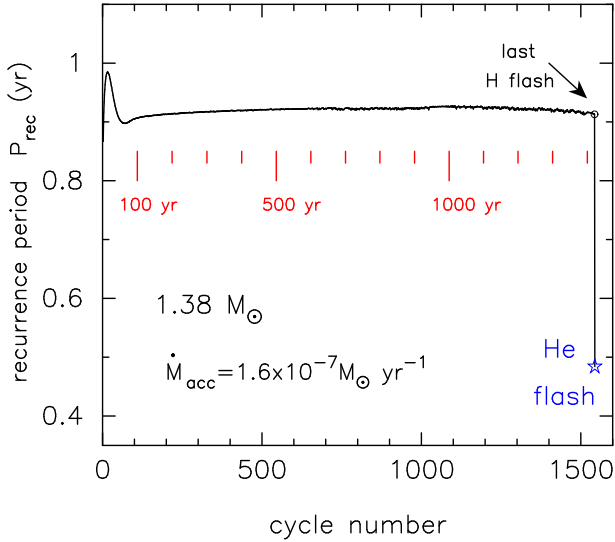}
\caption{
Change of the recurrence period 
on an accreting $1.38~M_\sun$ WD with a
mass-accretion rate of $1.6 \times 10^{-7}~M_\sun$~yr$^{-1}$. 
The abscissa shows the cycle number of hydrogen shell flashes. 
Time since the start of calculation 
is indicated every 100 years by the short vertical red lines. 
The positions of the final hydrogen shell flash (open black circle) 
and the He shell flash (open blue star) are also indicated. 
\label{periods}}
\end{figure}

\section{1500 Hydrogen Flashes}
\label{sec_Hflash}

\subsection{Nearly Identical Flashes with No Indication 
of Forthcoming Helium Flash}

Figure \ref{periods} shows the change of the recurrence period, 
$P_{\rm rec}$, throughout our calculation. 
After 1543 hydrogen flashes, a He flash occurs passing half a period 
from the last H flash (open blue star). 
The recurrence period increases by $\sim 8 \%$ 
just after the start of calculation
and then decreases during the first 70 cycles.  After that, it stays 
at $\sim 0.91$~yr. 
This early period change is caused by our choice of the initial WD model 
which is slightly different from the thermal equilibrium 
structure (see Section \ref{sec_initialmodel} for the effects 
owing to the choice of initial models). 


\begin{figure}
\epsscale{1.1}
\plotone{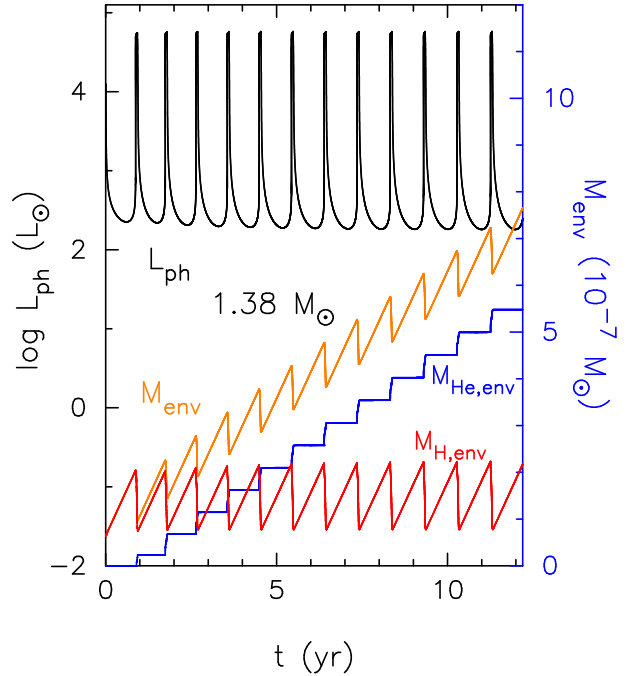}
\caption{The first 12 years of the evolution shown in Figure \ref{periods}. 
From top to bottom, the photospheric luminosity $L_{\rm ph}$ (black line), 
envelope mass, i.e., the total mass $- 1.38 ~M_\sun$ (orange line), 
helium envelope mass (blue line), and mass of hydrogen-rich envelope 
(red line). 
\label{12year}}
\end{figure}

Figure \ref{12year} shows the first 12 years of our calculation. 
There are 12 outbursts as shown by the change in the photospheric 
luminosity $L_{\rm ph}$. 
The mass of the hydrogen-rich envelope $M_{\rm H,env}$ 
(defined as the mass above $X > 0.01$) 
increases during the inter-pulse phase owing to accretion, and decreases 
during the outburst, owing partly to wind mass-loss 
and partly to hydrogen nuclear burning. The He envelope (defined as 
the region between the CO WD boundary and the bottom of a 
H-rich envelope) increases its mass 
during the outburst phase. 
The envelope mass, $M_{\rm env}$, which is the summation of the H-rich
and He envelopes, increases in the quiescent phases, and
sharply decreases due to wind mass-loss during the outburst phases.


\begin{figure}
\epsscale{1.1}
\plotone{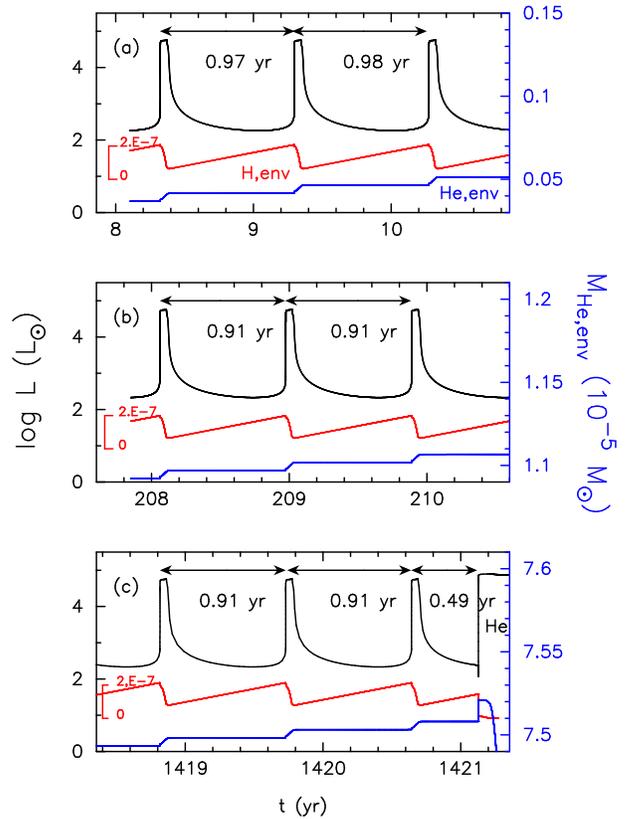}
\caption{
The evolution around (a) $t=10$ yr, (b) 210 yr, and (c) 1420 yr. 
The final hydrogen flash occurs at $t=1420.6$ yr 
followed by a helium flash. The hydrogen flashes around $t=210$ yr 
in panel (b) and 1420 yr in panel (c) are almost identical. 
\label{Levol}}
\end{figure}

Figure \ref{Levol} shows a close-up view of three flashes at 
three epochs of $t \sim$10 yr, $\sim$210 yr, and $\sim$1420 yr. 
As the recurrence period becomes constant after 70 years  
(see Figure \ref{periods}), the middle and bottom panels show 
the same $P_{\rm rec}=0.91$ yr. Through these flashes 
the He envelope steadily increases its mass with time. 
The bottom panel shows 
the last three outbursts before the He flash occurs
at $t=1421.13$ yr. 
Note that the photospheric luminosity $L_{\rm ph}$ and temperature  
$T_{\rm ph}$ change almost identically in panels (b) and (c), 
until the He nova outburst occurs. 
Thus, we have no precursors for the coming He nova outburst.


\begin{figure}
\epsscale{1.1}
\plotone{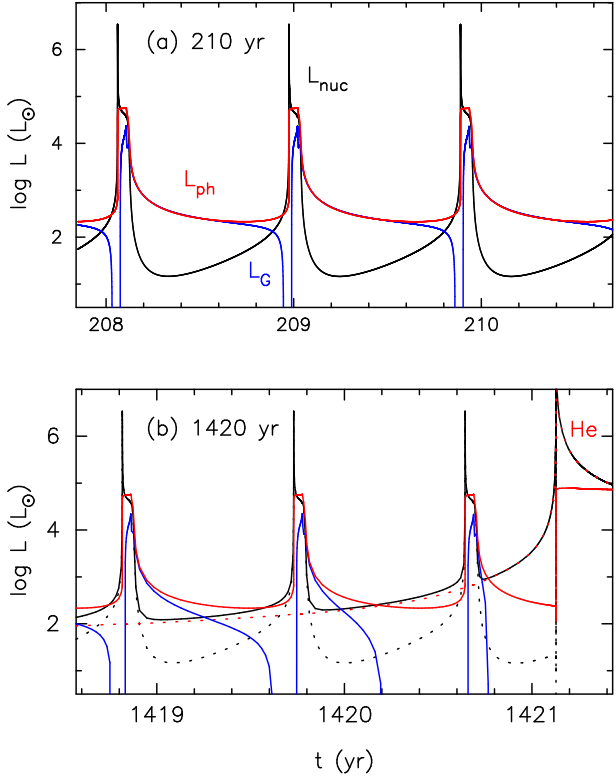}
\caption{
Temporal changes in the integrated fluxes of nuclear burning, 
$L_{\rm nuc}$ (solid black lines), photospheric luminosity, $L_{\rm ph}$
(solid red line), and integrated gravitational energy release, 
$L_{\rm G}$ (solid blue line). (a) Three flashes around $t=210$ yr. 
(b) Three flashes around $t= 1420$ yr just before the He flash. We also plot
the contributions of hydrogen burning (dotted black line) and
He burning (dotted red line) separately. 
\label{Levolnuc}}
\end{figure}

\subsection{Toward Helium Ignition} 

Although there are no apparent diagnostics of approaching a He nova 
in $L_{\rm ph}$ and $T_{\rm ph}$ until the He ignition, 
there is a gradual change in deep interior of the envelope. 
Figure \ref{Levolnuc} compares the energy budgets at 
$t\sim 210$ years and 1420 years. The photospheric luminosity, 
$L_{\rm ph}$, is the summation of the integrated nuclear burning 
rate, $L_{\rm nuc}$, and integrated gravitational energy release, 
$L_{\rm G}$. 
In the very early phase of the outburst, $L_{\rm nuc}$ reaches 
as large as $ > 10^6~L_\sun$, which is mostly absorbed 
in the burning region as indicated as $L_{\rm G} < 0$. 
As a result the photospheric luminosity is as small as 
$L_{\rm ph}\sim 5.8 \times 10^4~L_\sun$ at most in both the epochs. 

Figure \ref{Levolnuc} also shows that $L_{\rm G}$ turns 
from negative to positive after the peak of $L_{\rm nuc}$. 
The absorbed energy ($L_{\rm G} < 0$) is released in the later phase
of the outburst. 
This means that the burning region is slightly sinking back toward 
the original, geometrically thin,  plane parallel configuration. 
This energy release continues until the end of the H flash. 
The hydrogen-rich envelope mass has decreased owing to nuclear burning, 
being unable to support enough high temperature for hydrogen burning. 
Thus, nuclear burning extinguishes where  
$L_{\rm nuc}$ quickly decreases as shown in Figure \ref{Levolnuc}. 
In the interpulse phase, $L_{\rm G}$ owing to mass accretion
is the main source of radiation, $L_{\rm ph}$. 

Hydrogen nuclear burning is only the source of $L_{\rm nuc}$ at 
epoch $t\sim 210$ yr, while both H and He burning contribute 
to $L_{\rm nuc}$ at $t\sim 1420$ yr. 
We plot the contribution of hydrogen burning (dotted black line) and 
He burning (dotted red line) separately in Figure \ref{Levolnuc}(b).  
The hydrogen burning rate varies just in the same way as 
in $t\sim 210$ yr, but  
additional energy release owing to He burning continuously increases 
with time, which is absorbed in the inner envelope and 
not transferred upward.  
As a result, $L_{\rm ph}$
behaves just like the epoch of $t\sim 210$ yr. 

In this way, 
every hydrogen flash is almost identical until just before the He flash, 
in the recurrence period, flash duration, and quiescent luminosity,
even though the He envelope is growing in mass 
and its nuclear energy generation rate is increasing. 
In other words, there is no observational precursor 
of the forthcoming He flash. 

Note that helium ignited in the last interpulse phase of our H flash cycle 
calculation, i.e., a hydrogen flash itself does not trigger directly
the He ignition.  As shown later in Figure \ref{Trho.struc.He},
the temperature at the bottom of the He layer gradually increases
as its mass increases with time.  A small temperature peak appears
at the epoch of the last hydrogen flash, 
at $\log \rho ~({\rm g~cm}^{-3})\sim 5.1$, i.e., just above the CO core,  
where He burning already occurs with low rates. 
This small temperature peak eventually triggers the He flash 
during the interpulse phase.

\section{Helium Shell Flash}
\label{sec_Heflash}


\begin{figure}
\epsscale{1.1}
\plotone{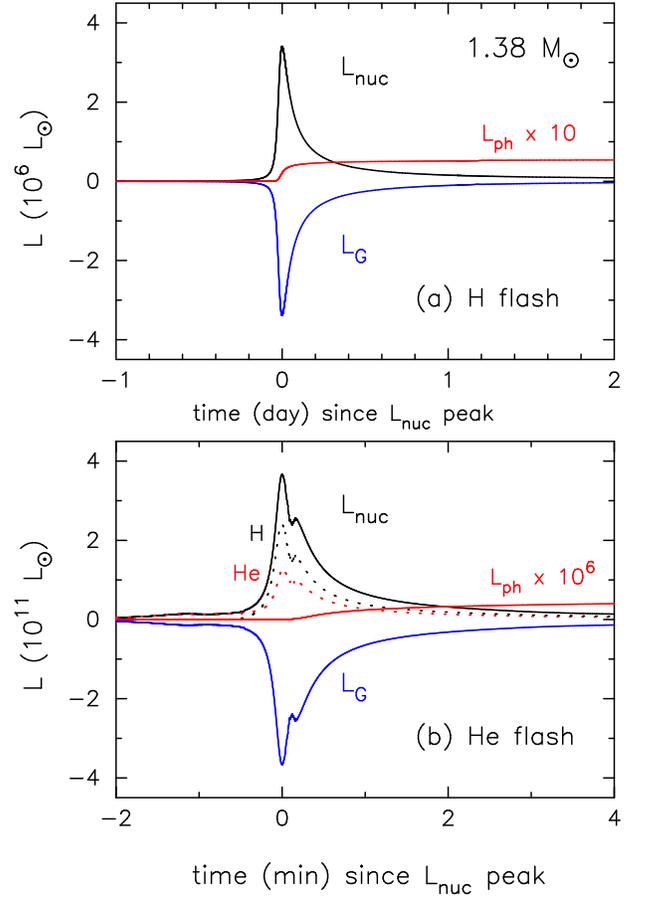}
\caption{Comparison of nuclear, gravitational, and photospheric 
luminosities at the ignition of the (a) H and (b) He shell 
flashes. The abscissae indicate the time since 
the integrated nuclear burning luminosity $L_{\rm nuc}$ (solid black lines)
reaches maximum. The solid blue lines indicate the integrated 
gravitational energy release rate, $L_{\rm G}$.
(a) The last H flash ($L_{\rm nuc}^{\rm max}$ is at $t=1420.64$~yr
in Figure \ref{Levolnuc}).
The solid red line indicates 10 times the photospheric luminosity, 
$L_{\rm ph} \times 10$. 
(b) The He flash ($L_{\rm nuc}^{\rm max}$ is at $t=1421.13$~yr). 
The solid red line indicates $10^6$ 
times the photospheric luminosity. 
The nuclear burning luminosity is divided into two parts: 
H burning (dotted black line) and He burning (dotted red line). 
\label{L.he}}
\end{figure}

\subsection{Onset of Helium Shell Flash}

Figure \ref{L.he} shows close-up views of very early phases of 
H and He shell flashes. 
The He flash is significantly different from the H flash in many points.
One of the differences is the nuclear energy release rate. 
At maximum, the integrated 
nuclear energy release rate reaches 
$L_{\rm nuc}^{\rm max}= 3.4\times 10^{6}~L_\odot$ in the H flash, 
whereas it reaches as large as 
$L_{\rm nuc}^{\rm max}= 3.7\times 10^{11}~L_\odot$ in the He flash,   
$10^5$ times larger than that of H burning. 
The timescale is also very different.  The H flash 
undergoes explosive nuclear burning in $\sim0.1$~days, 
whereas the He flash proceeds in as short as $\sim 0.5$~ minutes. 

In both the cases most of the nuclear energy is absorbed 
in the lower part of the burning region as 
shown by a large negative value of $L_{\rm G} (< 0)$.
Thus, only a very small part of the nuclear energy 
$L_{\rm nuc}$ is transported outward 
and emitted at the photosphere as $L_{\rm ph}$. 
As a result, the photospheric luminosity $L_{\rm ph}$
does not exceed the Eddington luminosity.
These properties were already reported for the H flash 
model of $1.38~M_\sun$ WD \citep[Figure 2 in][]{kat16xflash}. 
The present calculation demonstrates that 
the He nova has similar properties 
even for much larger nuclear energy release rates
and much shorter timescales.


\begin{figure*}
\epsscale{0.75}
\plotone{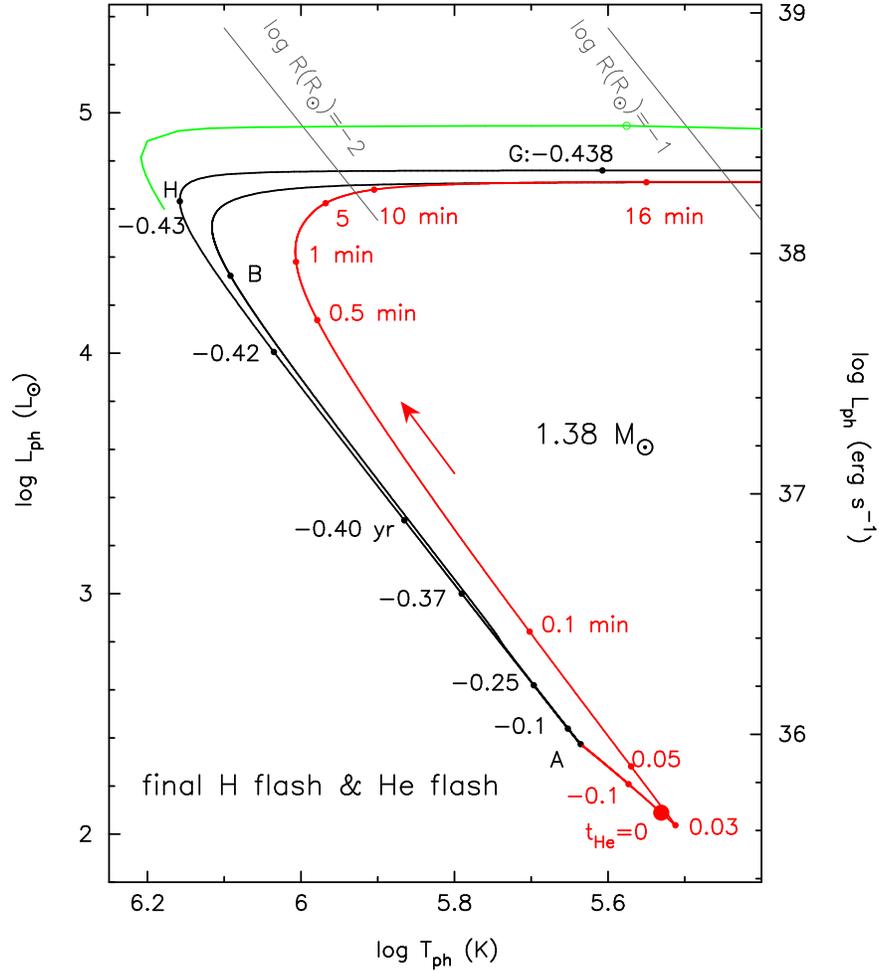}
\caption{Evolutional path in the H-R diagram from $t=1420.4$ yr 
to 1421.2 yr, 
which corresponds to the evolution presented 
in Figures \ref{Levol}(c) and \ref{Levolnuc}(b). 
The solid black line shows the evolution in the final H flash
at $t=1420.64$ yr (stages A-B-G-H and later), 
and the red line shows the rising phase of the He flash. 
The marked stages correspond to:
A, interpulse phase before the final H flash occurs, 
i.e., $L_{\rm ph}^{\rm min}$ ($t=1420.40$ yr);
B, the epoch at the maximum integrated nuclear burning rate, 
$L_{\rm nuc}^{\rm max}$ ($t=1420.64$ yr), of the final H flash; 
H, $T_{\rm ph}^{\rm max}$ ($t=1420.69$ yr). 
The other stages are indicated by the time from 
the onset of He flash, i.e., $L_{\rm nuc}^{\rm max}$  
at $t=1421.13$ yr (marked by the large filled red circle). 
The black and red numbers along the evolutionary track indicate 
$t_{\rm He}$ in units of years and in units of minutes, respectively.  
The green curve denotes a steady-state sequence of a He nova that 
represent the decay phase of a He flash (see Section \ref{sec_heflash} 
for more details). 
\label{hr}}
\end{figure*}

\subsection{H-R Diagram and X-ray Flash}
\label{sec_hr}

Figure \ref{hr} shows the track in the H-R diagram for 
the final H flash (solid black line) 
followed by the He shell flash (solid red line). 
Now we define the time $t_{\rm He}$ starting at 
the onset of the He flash. We set $t_{\rm He}=0$ 
when the total nuclear energy generation rate 
reaches its maximum, $L_{\rm nuc}=L_{\rm nuc}^{\rm max}$.  
After the final hydrogen flash, 
the star becomes faint keeping the photospheric radius 
almost constant, from point H (maximum $T_{\rm ph}$) toward the point 
of $t_{\rm He}=0.0$.  It takes 0.43 yr.  After that the star 
brightens up within a minute, 
along with a constant but a bit larger than the
photospheric radius in the case of H flashes.  

A similar track in the H-R diagram was already shown by \citet{ibe83} 
as an evolution passing the phase of a planetary nebula nucleus of 
an $0.6~M_\sun$ 
post-AGB star. As the star evolves down in the H-R diagram, 
a final He flash occurs and the star brightens up again. 
The decay timescale of the final H shell flash is as long as 4000 yr, 
while the rising timescale of the He shell flash is 20 yr. 
These values should not be
directly compared with our case because the WD mass is different, 
but we see similar characteristic properties of a He shell flash. 


\begin{figure}
\epsscale{1.1}
\plotone{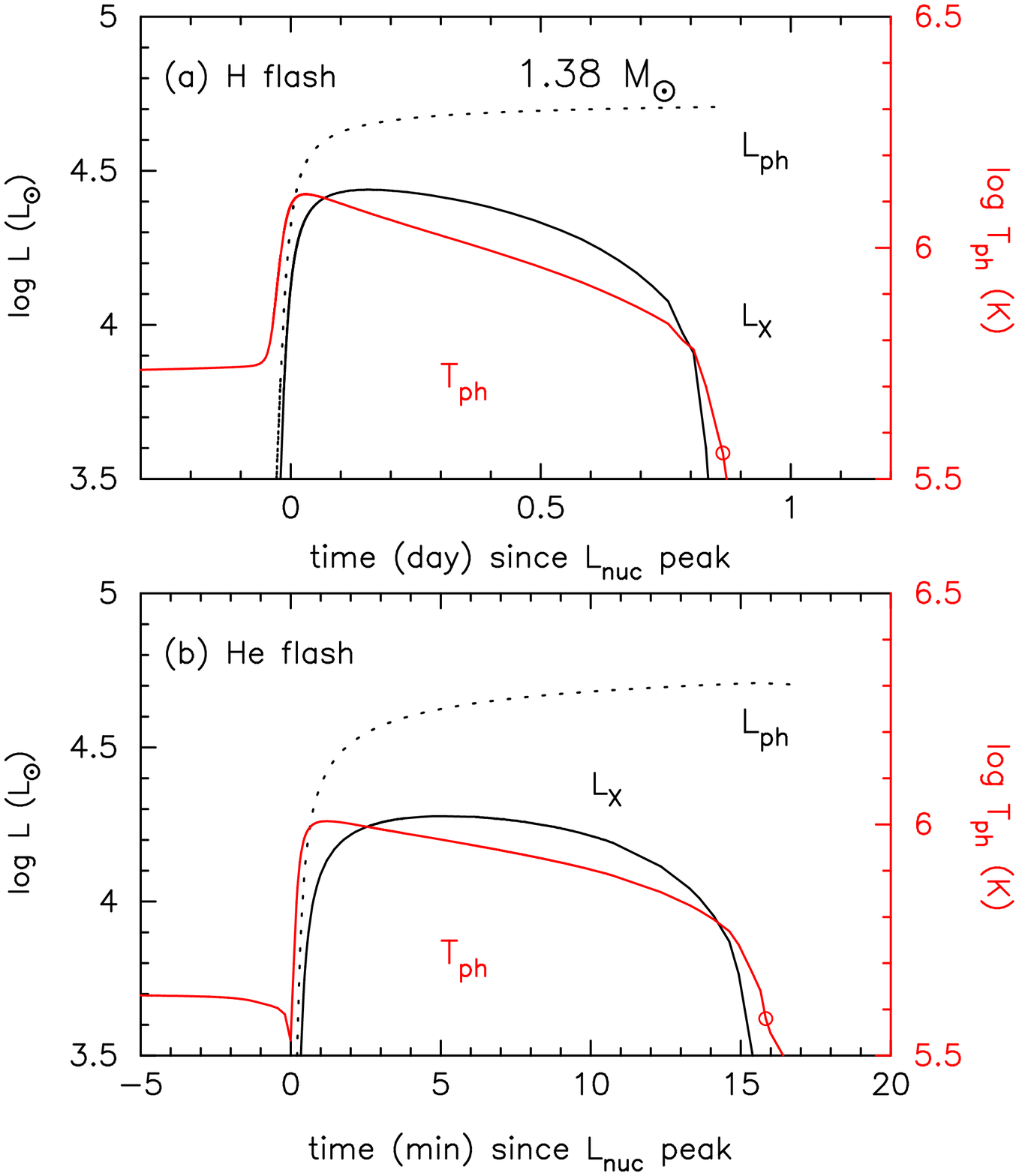}
\caption{Comparison of the prompt X-ray flash phases in the (a) H 
and (b) He flashes.  
Solid black lines represent $L_{\rm X}$, the X-ray luminosity 
in the (0.3 -- 1.0)~keV band.  Dotted black lines: $L_{\rm ph}$. 
Solid Red lines:  $T_{\rm ph}$. 
The open circles denote the stage when the optically thick winds occur.   
\label{xflash}}
\end{figure}

Figure \ref{xflash} demonstrates the difference in the rising timescales 
between H and He flashes. 
The figure shows the evolutions of the photospheric temperature, 
total luminosity, and (0.3 -- 1.0) keV X-ray luminosity for the H 
and He flashes. 
The X-ray luminosities are calculated assuming the black body spectrum 
of the photospheric temperature. This assumption may not be 
accurate for observational X-ray fluxes but enough to estimate 
the duration of the X-ray flash, because 
the rising and decay timescales are very short compared with the duration.
The X-ray flash is a brief X-ray bright phase 
in the very early phase of the outburst before the optical maximum 
\citep[see, e.g.,][]{kat16xflash}. 
As the He flash is much more violent, the duration of the X-ray flash is 
as short as 15 min and much shorter than that of the H flash 
($\sim 1$~day).   Nevertheless, the X-ray luminosity is a little bit
smaller than that of the H flash because of a lower
photospheric temperature (red line in Figure \ref{hr}). 
Thus, detection of X-ray flash in the He flash would be difficult even 
in high cadence satellite observations as planned for the X-ray flash 
in the 2015 outburst of M31N 2008-12a \citep{kat16xflash}. 


\begin{figure}
\epsscale{1.1}
\plotone{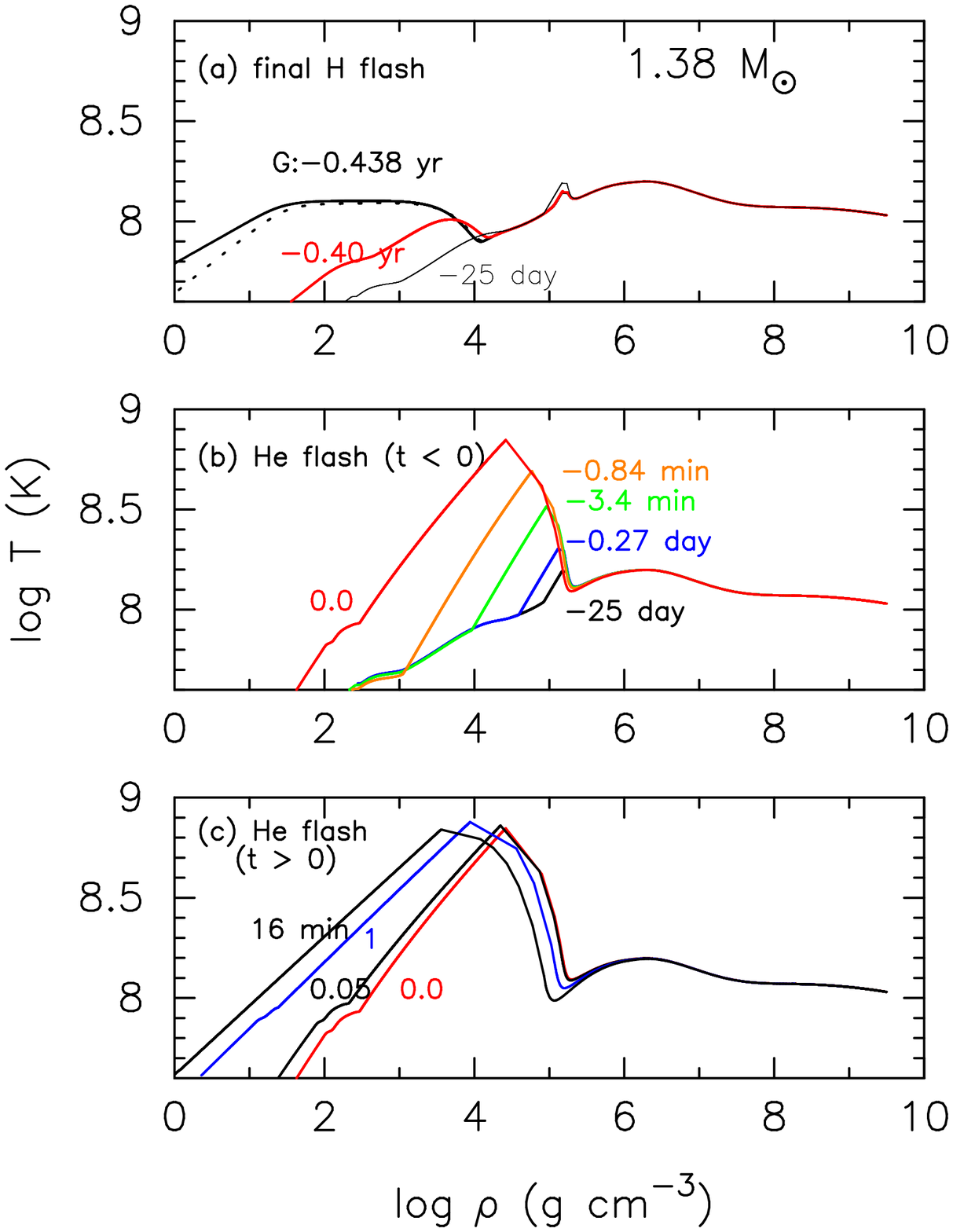}
\caption{Temporal change of the envelope structure in the $\rho$-$T$ plane
for the final H flash and the He flash.
(a) Structures for selected stages in the final H flash: 
stage G (thick solid black line), H (dotted black line), 
$t_{\rm He}=-0.40$ yr (solid red line) and $-25$ days that locates between stages 
$-0.1$ yr and A in Figure \ref{hr} (thin solid black line).   
The other two panels represent the evolution in the He flash:
(b) before ($t_{\rm He} < 0$) and (c) after ($t_{\rm He} > 0$) 
the time of $L_{\rm nuc}^{\rm max}$.
\label{Trho.struc.He}}
\end{figure}

\subsection{Internal Structures before/after Helium Ignition}
\label{sec_structure}

Figure \ref{Trho.struc.He} shows the temporal changes 
of the internal structure 
in the $\rho$-$T$ plane during the course of H and He shell flashes. 
The rightmost point corresponds to the center of the WD. 
Figure \ref{Trho.struc.He}(a) shows the structure change 
in the later phase of the final H flash, starting from stage G 
in Figure \ref{hr}, 
which roughly corresponds to the beginning of a late supersoft
X-ray source (SSS) phase. 
Hydrogen ignites at the bottom of the H-rich envelope, i.e., 
$\log \rho ~({\rm g~cm}^{-3})\sim 2$, but until this stage, heat was transferred  
both outward and inward to form a large hot region 
($0< \log \rho~({\rm g~cm}^{-3}) < 4$). 
The internal structure hardly changes   
in the SSS phase (from stage G to stage H: dotted line).  
In the following cooling phase toward $t_{\rm He}=-25$ days, 
the star moves down in the H-R diagram (Figure \ref{hr}) 
from stage H to stage A along with a constant radius.  

The temperature profile ($\log \rho~({\rm g~cm}^{-3}) < 4$)
changes up and down, in the flash and interpulse phases, in every cycle. 
On the other hand, the temperature in the deep interior of the envelope 
hardly changes, 
except that a tiny peak appears at the base of 
the He zone ($\log \rho ~({\rm g~cm}^{-3})= 5.2$) in the last three H flashes. 

The middle and bottom panels show the temperature change immediately 
before and after the He ignition ($t_{\rm He} = 0$), respectively. 
The middle panel demonstrates that the 
tiny peak at $\log \rho~({\rm g~cm}^{-3})= 5.2$ in the top panel 
extends toward lower-density region and the maximum temperature increases 
to $\log T$ (K) $\sim 8.8$. 
A convection zone develops outward from the temperature maximum. 
Lower density layers 
become hotter and hotter, whereas   
the temperature at the He burning zone gradually drops because of
adiabatic expansion (negative $L_{\rm G}$). 


\begin{figure}
\epsscale{1.15}
\plotone{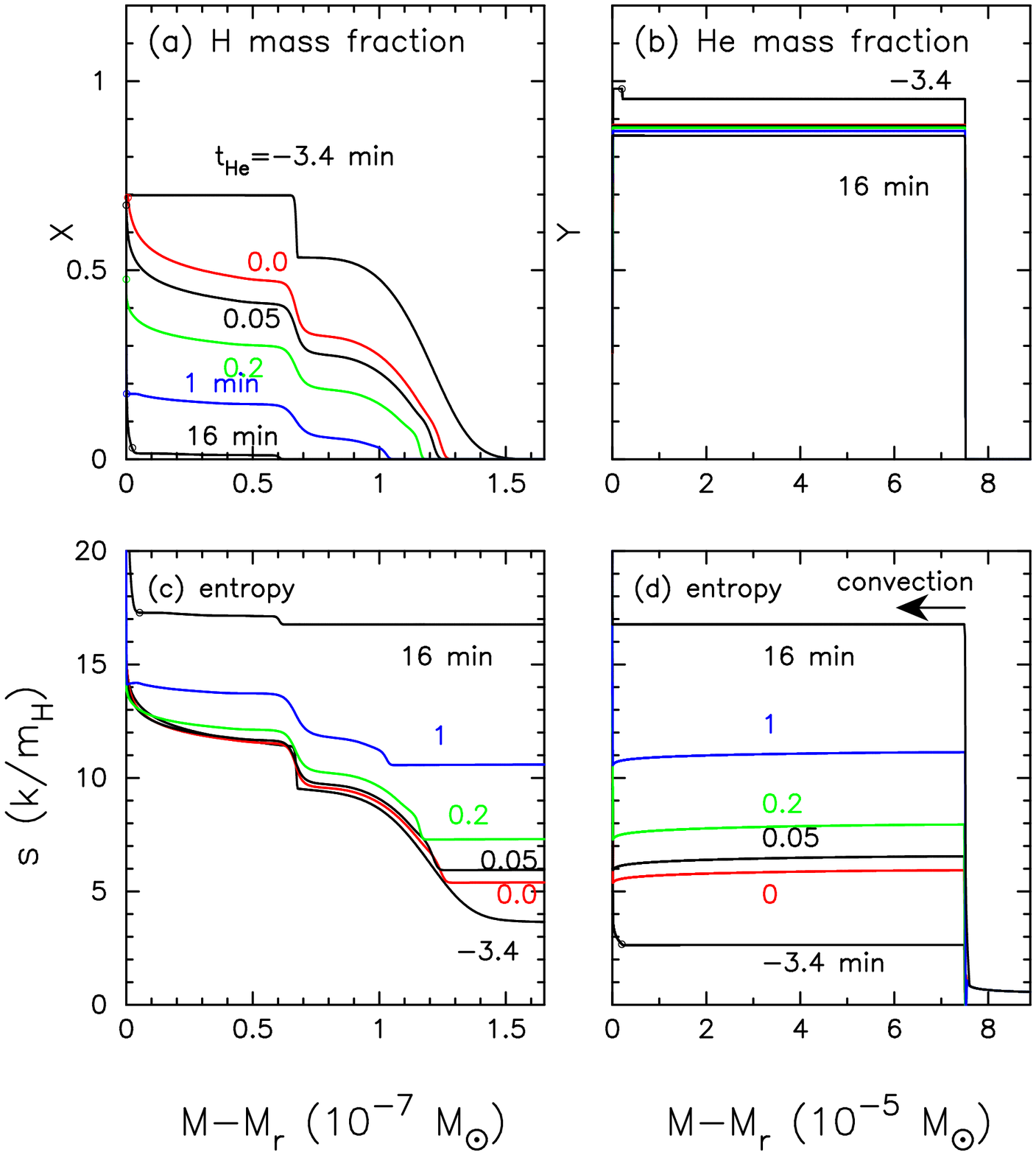}
\caption{Hydrogen and helium mass fractions, and entropy distributions,
in the envelope before/after the He ignition. The time from the He ignition
($L_{\rm nuc}^{\rm max}$) is indicated beside each line. The 
same color indicates the same stage in other panels. 
(a) Temporal change of hydrogen mass fraction in 
the uppermost layer. 
(b) Helium mass fraction in a more wider region down to the upper part 
of the CO core. The boundary between the envelope and CO core is at 
$M-M_r=7.5\times 10^{-5}~M_\sun$. 
(c) Entropy distribution in the same region of panel (a), 
 in units of $k/m_{\rm H}$, where
$k$ is the Boltzmann constant and $m_{\rm H}$ is the mean atomic mass.  
(d) Entropy distribution in the same region of panel (b). 
Convection occurs in all the stages.  The bottom of the convective region 
is indicated by the origin of the leftward arrow in panel (d).  
We indicate the outer edge of convective region by small open circles.
At $t_{\rm He}=-3.4$ min, the convection does not yet reach 
the region of panel (a), so the edge is indicated in panels (b) and (d). 
In the later stages, the convection widely
develops up to the H-rich envelope and the outer edge is indicated in 
panel (a), and in panel (c) (only for $t_{\rm He}=16$ min).  
The optically thick wind mass-loss occurs just after the stage $t_{\rm He}=16$ min. 
\label{qxs}}
\end{figure}

\subsection{Mixing of Hydrogen into Helium Burning Zone}
\label{sec_Hmixing}

Figures \ref{qxs}(a) and (b) show the temporal change 
of the H/He profile, while Figures \ref{qxs}(c) and (d) show 
the entropy distribution in the corresponding stages.  
After the final H-flash, freshly accreted matter 
($X=0.7$, $Y=0.28$, and $Z=0.02$) accumulates 
on top of the H-rich/He layer. 
Figure \ref{qxs}(a) shows that, at $t_{\rm He}=-3.4$ min, the freshly accreted
layer of $X=0.7$ lies on top of 
the leftover of the final H-flash, where $X$ is decreasing inward. 

Convection occurs, before $t_{\rm He}=0$, at the He nuclear 
burning region.  The convection spreads almost all over the envelope. 
The inner edge of the convective region is shown as the origin 
of the black arrow in Figure \ref{qxs}(d). 
The outer edge of the convective region 
is indicated by the small open circles.  
Note that the convection does not reach the photosphere, 
thus, the surface hydrogen content is always $X=0.7$ until the wind occurs 
at $t_{\rm He}=16.1$ min.

The convective region extends all over the He layer and 
penetrates into the upper H-rich envelope 
(see Figures \ref{qxs}(c) and (d)).  
The H-rich matter is carried inward and mixed into deep interior  
of He-rich zone where the temperature is very high, 
and hydrogen is burned into helium. 
Therefore, the H mass fraction rapidly decreases with time. 
As shown in panel (a), most hydrogen disappears 
until $t_{\rm He}=16$ min.


\begin{figure}
\epsscale{1.08}
\plotone{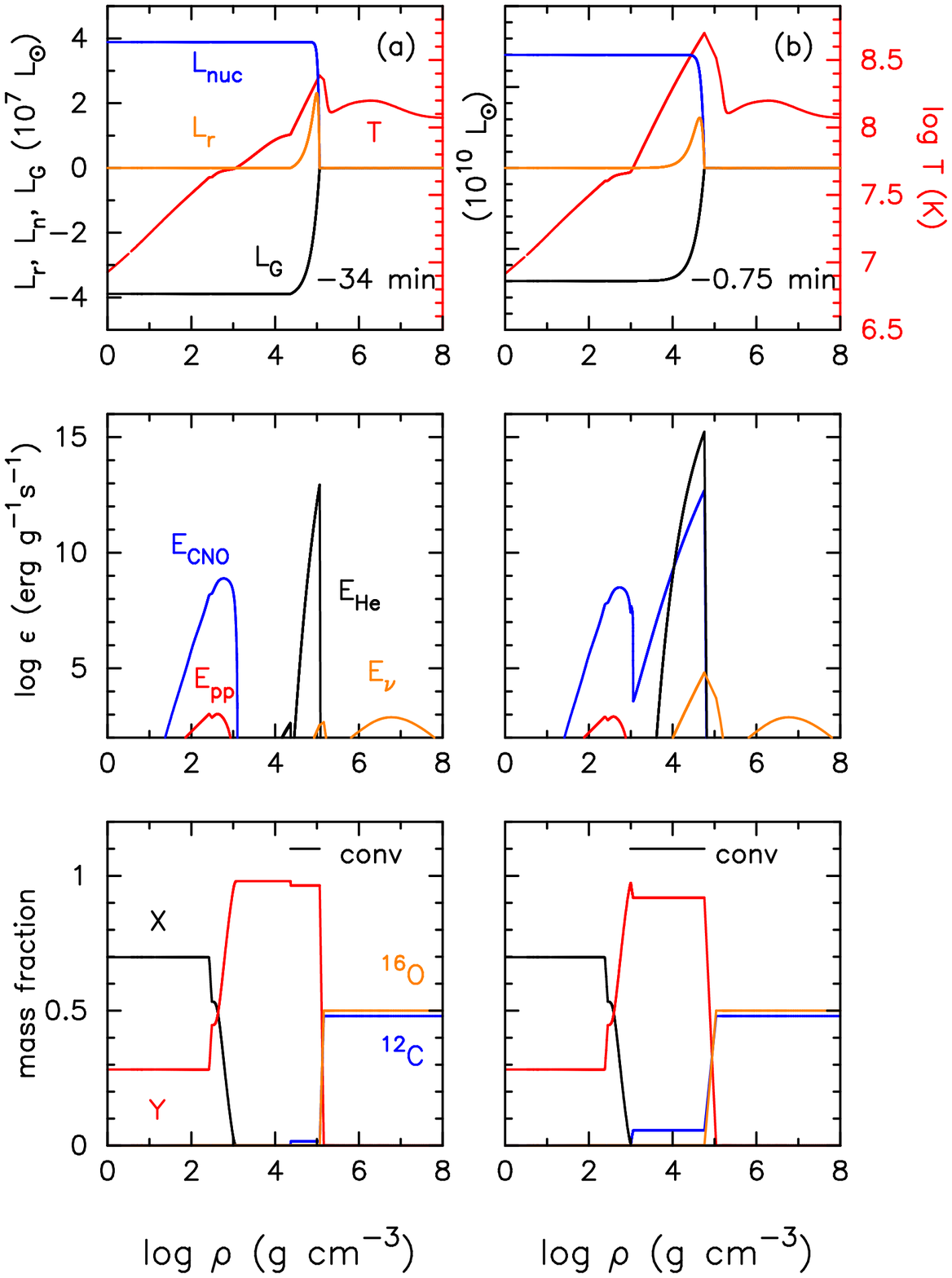}
\caption{Temporal change of physical quantities representing nuclear burning. 
(a) $t_{\rm He}=-34$ min, (b) $-0.75$ min.  
Upper panels: $L_{\rm nuc}$, $L_{\rm G}$, and $L_r$,  
where $L_{\rm nuc}$ is the summation of H and He burning. 
The temperature profile is also added (red line).
Middle panels: 
the specific energy generation rate owing to {\it pp}-chain (red), 
CNO-cycle (blue), He burning (black), and neutrino energy loss (orange). 
Bottom panels: the mass fraction of hydrogen ($X$, black), helium 
($Y$, red), carbon ($^{12}$C, blue), and oxygen ($^{16}$O, orange). 
The horizontal bar indicates the convective region at each stage.  
\label{ulg.beforeHe}}
\end{figure}

\begin{figure}
\epsscale{1.15}
\plotone{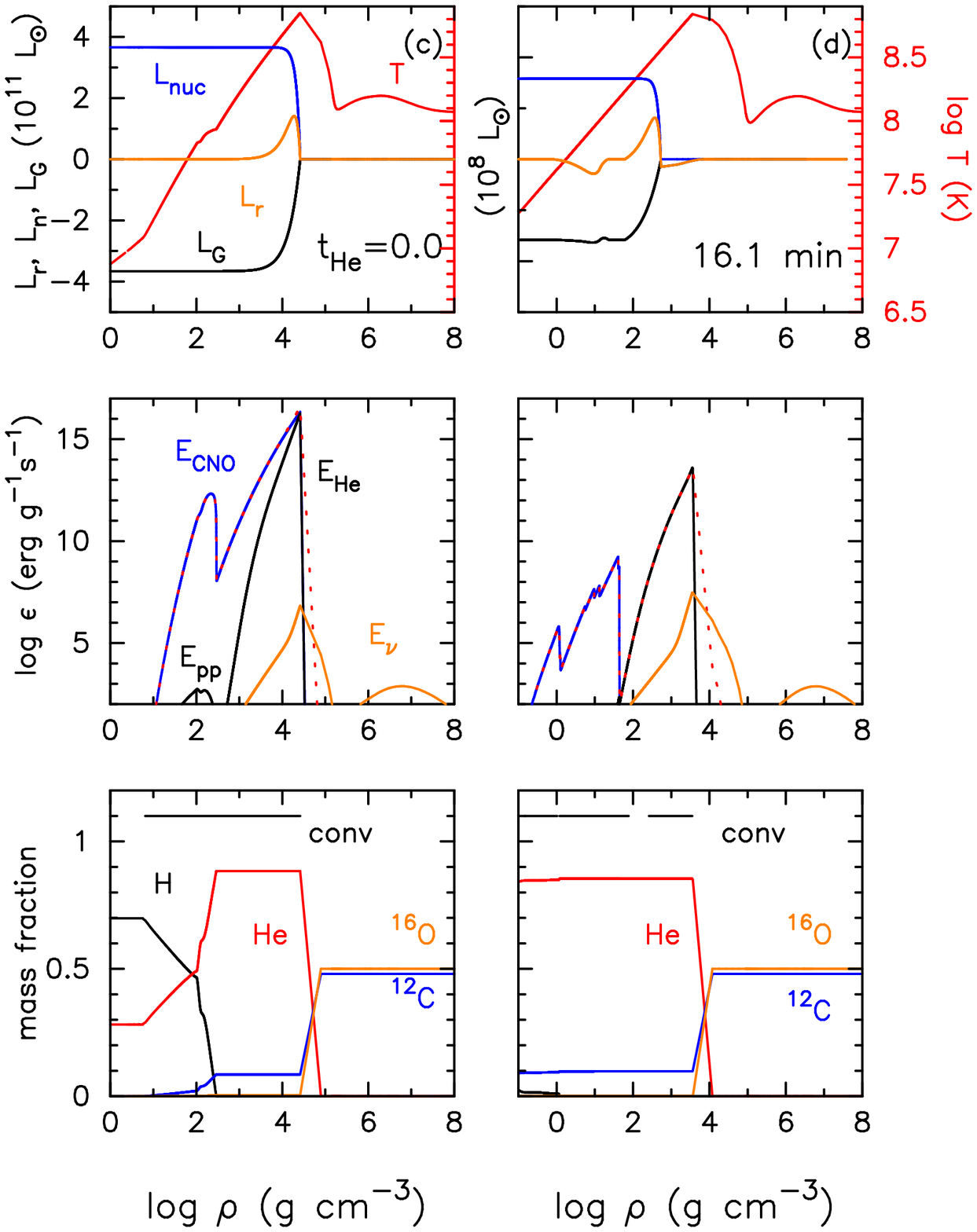}
\caption{Continuation of Figure \ref{ulg.beforeHe}, but for 
(c) $t_{\rm He} = 0.0$ min, and (d) $t_{\rm He}=16.1$ min 
(onset of the wind mass-loss).  At this epoch,
convection separately occurs in the two regions 
corresponding to H and He burning. 
Middle panels: the dashed red lines indicate the total nuclear burning 
rate, i.e., the difference to $E_{\rm He}$ indicates 
nuclear reactions of heavier elements other than triple alpha reaction at 
the boundary between the CO core and He-rich envelope.  
\label{ulg.afterHe}}
\end{figure}

Figures \ref{ulg.beforeHe} and \ref{ulg.afterHe} show the changes of 
the energy budget, nuclear-burning energy-generation rate, and chemical 
composition, for the selected stages of (a) $t_{\rm He}=-34$ min, 
(b) $-0.75$ min, (c) 0.0 min, and (d) 16.1 min. 
The top panels show 
the nuclear luminosity integrated from the center of the WD up to the 
radius $r$, $L_{\rm nuc}(r)$ (solid blue lines),   
integrated gravitational energy release rate, $L_{\rm G}(r)$ 
(solid black lines), and local luminosity at radius 
$r$, $L_{r}$ (solid orange lines),  
which is the sum of radiative and convective luminosities. 
The temperature profile (red lines) is added.  
The middle panels show the energy generation rates per unit mass 
owing to {\it pp}-chain, CNO-cycle, and He burning, 
and the neutrino energy loss rate.
The bottom panels show the mass fractions of selected elements. 

The convective region is also indicated by the short horizontal bars 
in the bottom panels.  
The convection started from the He burning region and spreads 
outward all over the He layer and penetrates into the H-rich envelope 
until $t_{\rm He}=0$.
Thus, the surface hydrogen is mixed into the inner He layer 
and explosively burns with very high temperatures. 
As shown in the middle panels,  
the specific energy generation rate owing to CNO-cycle becomes comparable 
to that of He burning even in the region where the H mass fraction 
is very small. 
Therefore, hydrogen burning contributes as much as two thirds 
to the total nuclear burning rate $L_{\rm nuc}$ as in
Figure \ref{L.he}(b).

\begin{figure}
\epsscale{1.15}
\plotone{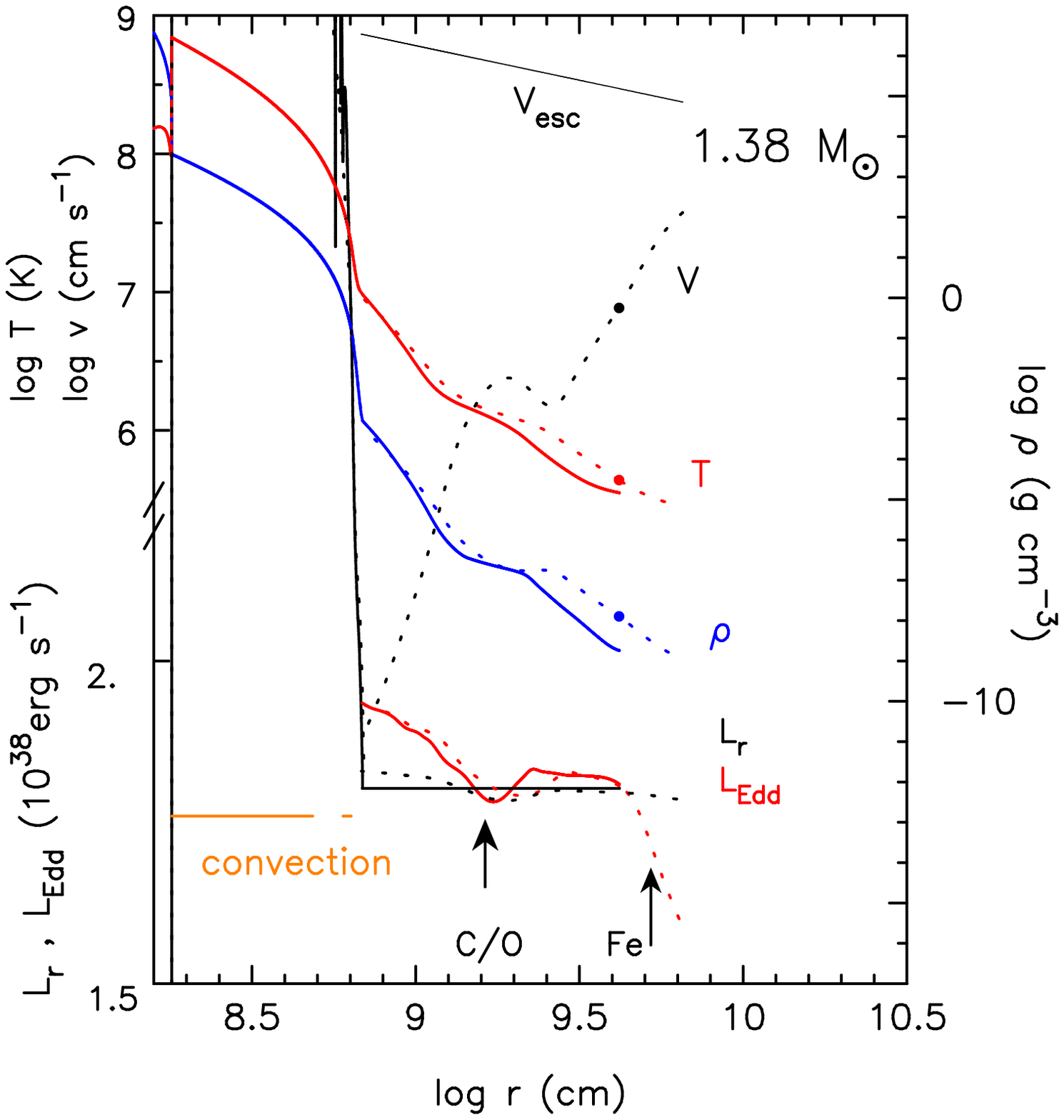}
\caption{Envelope structures just before (solid lines:  
$\log T_{\rm ph}$ (K) = 5.55 in Figure \ref{hr}) 
and shortly after (dotted lines:  
$\log T_{\rm ph}$ (K) = 5.45) the optically thick 
winds occur in the He shell flash. 
From upper to lower, the escape velocity 
$V_{\rm esc}=\sqrt{2GM_{\rm WD}/r}$,
wind velocity $V$ for the wind model, 
temperature $T$ (upper red lines), 
density $\rho$ (blue lines), 
local luminosity $L_r$ which is the summation of radiative luminosity 
and convective luminosity (lower black lines), 
and the local Eddington luminosity
$L_{\rm Edd}=4 \pi c G M_{\rm WD}/ \kappa$ (lower red lines).
The position of the sonic point \citep[critical point,][]{kat94h}
is indicated by filled circles.
The outermost point of each structure line corresponds to the 
photosphere. 
The two black arrows indicate the inner edges of 
the opacity peaks owing to C/O and Fe. 
 The convective regions are indicated by the horizontal orange lines.
\label{struc.startML}}
\end{figure}

\subsection{Occurrence of Optically Thick Wind}

Figure \ref{struc.startML} shows 
envelope structures just before and after the optically thick
winds occur.  The solid/dashed lines denote the structures 
just before/after the optically thick winds occur.
We also indicate two places corresponding to the inner edge of the opacity
peak owing to highly ionized Fe, C, O, and Ne (labeled ``C/O'') 
and inner edge of the peak owing to low/mid-degree ionized iron
(labeled ``Fe'') by the arrows. 
The optically thick winds are driven by the Fe opacity peak, 
not by the C/O peak, because the sonic point \citep[critical point,][]{kat94h}
of the optically thick winds is located
at the inner edge of the Fe opacity peak, not at the C/O peak.  
The structure changes little at the onset of optically thick winds. 
This property is the same as that of acceleration in hydrogen flashes 
\citep{kat16xflash}.

Our calculation of the He flash stopped at $t_{\rm He}=1.6$ years
before it evolves to a SSS phase because of numerical difficulties. 
As well known Henyey-type code calculations do not work
(do not converge) 
when the envelope extends to a giant size and surface region becomes 
radiation-pressure dominant. One way to continue numerical calculation 
is to assume very large mass-loss rates, but such large rates are often 
inconsistent with realistic wind acceleration such as optically-thick winds. 
In our previous paper, we developed an iteration method to 
calculate a complete cycle of hydrogen shell flashes with 
mass-loss rates consistent with the optically-thick winds 
\citep{kat17sh}.
However, the He shell flash is so violent that we did not succeed in
calculating the wind mass-loss phase with the iteration method. Thus, 
in this work, we did not adopt the iteration cycle,  
instead, we assumed relatively large trial mass-loss rates 
during the He flash. 
Although we assumed mass-loss rates as small as possible, they are 
much larger than those of the optically-thick winds. This makes 
the outburst evolution faster, so we sickly underestimate 
the flash duration.
We suppose the duration of the He flash possibly longer 
than 1.6 years, namely about 2 years or more.

\section{Discussion}
\label{sec_discussion}

\subsection{Initial WD Model and Its Central Temperature}
\label{sec_initialmodel}

\citet{pri95} and \citet{yar05} presented nova calculations 
for a wide range of three parameters,  
the WD mass, central temperature of the WD, and mass-accretion rate. 
These three parameters are not independent of each other 
but linked through long-term evolution of the binary system. 
\citet{epe07} calculated 1000 cycles of H flashes on a 1.0 $M_\sun$ WD 
with a mass-accretion rate of $1\times 10^{-11}M_\sun$~yr$^{-1}$. 
The recurrence period quickly increases by a factor of 10 from the initial 
$1.8\times 10^6$ years in the first 400 cycles
and then turn to a gradual increase to $2.18\times 10^7$ years in the 
final 100 cycles (see their Figure 2). 
They adopted an initial WD temperature of $T_{\rm c}=3 \times 10^7$ K, and  
during the calculation, the central temperature decreases with time by 
a factor of 5.  Such a large change occurs because they assumed 
a much hotter initial WD model than that of 
an equilibrium model with the corresponding mass-accretion rate. 
The above authors took a 0.6 $M_\sun$ WD with 
$\dot M_{\rm acc}=1.0 \times 10^{-9}M_\sun$~yr$^{-1}$ and 
showed that an initially hotter WD of 
$5 \times 10^7$ K cools down but an initially cooler WD of 
$5 \times 10^6$ K becomes hotter and the WD temperatures approach
a common equilibrium value after 3000 cycles. 
This means that if they adopted an initial WD model close to that of 
the equilibrium model, nova cycle would approach a steady-state 
much earlier. 

\citet{hil16} also presented a similar phenomena, 
in successive helium shell flashes in a He accreting WD 
with a mass-accretion rate of $2.0 \times 10^{-7}M_\sun$~yr$^{-1}$. 
The central temperature increased 
by a factor of 4 through 400 helium shell flashes 
during which the WD mass increases from 1.105 $M_\sun$ to 1.247 $M_\sun$.

In the present paper we adopted an initial WD model very close to the 
thermal equilibrium with the mass-accretion rate 
(see Section \ref{sec_method}). 
Thus, the WD interior is already hot and the central
temperature is as high as $\log T$ (K)=8.0299 at the start of 
calculation ($t=0.0$ yr)
that is close to the final value 
of $\log T$ (K)=8.0304 at the onset of He shell flash ($t=1421.13$ yr). 
Therefore, in our calculation, the recurrence period soon (after 70 cycles) 
approaches the final value with a small amplitude (8 \%)
variation (see Figure \ref{periods}).

\citet{kat15sh, kat17sh} suggested that 
the recurrent nova M31N 2008-12a is consistent with 
the 1.38 $M_\sun$ WD model because of its short recurrence 
period and rapid decline. 
Such an extremely massive WD is unlikely born as it is, 
but likely has grown up through long-term mass-accretion  
from the companion star \citep[see, e.g.,][]{hkn99, hknu99}. 
Thus, the WD is likely as hot as expected in an equilibrium model 
with the mass-accretion rate of $\sim 10^{-7}M_\sun$~yr$^{-1}$.  
Therefore, our assumption of the initially hot WD is reasonable.

\subsection{Accretion Energy} \label{sec_accenergy}

We suppose that gas is accreted through an accretion disk, releasing 
a part of gravitational energy from its surface which is 
emitted perpendicularly to the disk. 
Still, remaining energy is expected to be released in the boundary 
layer. We have neglected the energy released above the photosphere 
as in other previous nova calculations 
\citep[e.g.][]{ida13,wol13a,hac16sk}, whereas 
\citet{pri95} included the energy from boundary layer of which 
amounts 15 \% of the gravitational energy based on the work by \citet{reg89}. 
The 15 \% of the gravitational energy release rate corresponds to 
$1.5 \times10^{36}$~erg~s$^{-1}$ ($390~L_\sun$) in our case.  
Thus, the quiescent luminosity increases to $\log L$ (erg~s$^{-1}) =36.37$. 
The additional energy, however, hardly causes appreciable effects 
in the nova calculations as discussed below.

The heat flux from the boundary layer amounts 1.5 times the quiescent 
phase luminosity of our model. 
If this additional heat source changes thermal structure deep interior, 
the flash properties, such as the ignition mass,  
maximum temperature, and recurrence period may change. 
\citet{pri95} included 15 \% of the gravitational energy, and their 
$1.0~M_\sun$ WD model with the mass accretion rate of 
$1.0 \times 10^{-6}~M_\odot$~yr$^{-1}$ ($T_{\rm WD}=5 \times 10^7$ K) 
has the accreted mass of $2.15 \times 10^{-6}~M_\odot$ and 
the maximum temperature $T_{\rm max}=1.03 \times 10^8$ K.   
For the same WD mass and accretion rate,  
\citet{hac16sk} obtained $2.06 \times 10^{-6}~M_\odot$ and 
$T_{\rm max}=1.06 \times 10^8$ K. 
Considering the differences in the input parameters,  
these two models are in good agreement. 
\citet{hac16sk} also obtained a 1.35 $M_\sun$ model with 
$\dot M_{\rm acc}=5 \times 10^{-7}M_\sun$~yr$^{-1}$ that shows 
$2.0 \times 10^{-7}M_\sun$ and $T_{\rm max}=1.47 \times 10^8$ K, 
also being consistent with the grid models of 
1.25 $M_\sun$ and 1.4 $M_\sun$ with 
$\dot M_{\rm acc}=1 \times 10^{-7}M_\sun$~yr$^{-1}$  and 
$\dot M_{\rm acc}=1 \times 10^{-6}M_\sun$~yr$^{-1}$ in \citet{pri95}. 
Thus, we may conclude that the inclusion of the additional 
15 \% of the gravitational energy release 
does not make much difference on the flash properties.

\subsection{Mass Accumulation Efficiency}
\label{section_assumption}

In our long-term evolution model, a part of the accreted H-rich matter 
is lost during the wind phase. 
In the present work, we adopted slightly overestimated mass-loss rates 
as described in Section \ref{sec_method}, to simplify our calculation. 
The adopted wind mass-loss rates result in the 64\% lost of mass
during one cycle of H flash. 
Thus, the mass accumulation efficiency, $\eta\equiv 1-$
(the ratio of lost mass to accreted mass), 
is $\eta=0.36$.  This ratio increases to $\eta=0.40$
if we use the self-consistent wind mass-loss rates \citep{kat17sh}.

This accumulation efficiency is, however, highly uncertain 
because there are many unsolved problems associated with nova light curves. 
In classical novae, theoretical free-free emission light curves 
well reproduce the decay phase of light curves in optical 
and NIR bands \citep[i.e.,][]{hac06kb,hac10k,hac15k,hac16k}. 
We expect that the optical peak corresponds to the peak of 
the wind mass-loss rate, because the free-free emission optical flux is 
in proportion to the square of the wind mass-loss rate. 

In the rising phase, on the other hand, no reliable light curves 
have been calculated neither for classical novae nor for recurrent novae. 
\citet{hac14} presented an idea for the pre-maximum evolution, 
based on color-color evolution of fast novae, that 
the mass ejection begins just before the optical maximum. 
The quick start of the mass ejection in our model may correspond to 
the fast expansion of the photosphere rising toward optical/NIR maximum. 
 
If this is the case, the real mass-ejection begins shortly before the 
optical peak. In M31N 2008-12a outbursts, optical/NIR magnitudes 
rose in a short time toward the peak ($< 1$ day) 
\citep[e.g.][]{dar16}. This timescale is much 
shorter than the slow pre-maximum evolution 
of theoretical model \citep{kat16xflash,kat17sh}, in which the 
wind phase lasts 2 weeks before the peak.  
Moreover, the wind velocity does not reach the escape velocity 
$V_{\rm esc}=\sqrt{2GM_{\rm WD}/r_{\rm ph}}$
in the beginning of the wind phase \citep[see Figure 7 of][]{kat16xflash}. 

One possible idea to solve the inconsistency is that, 
in the premaximum phase, the wind solutions should be treated as 
a theoretical representative of fast-expanding surface
of the hydrogen-rich envelope when the static solutions do not exist 
as the luminosity approaches the Eddington luminosity, 
and the wind mass-loss rate is just a parameter to characterize 
the expanding envelope solutions. 
In this case, wind solutions give the photospheric temperature, 
radius, and luminosity, but the wind mass-loss rates 
should not be taken as real mass-outflow rates. 
If we adopt this idea, the mass lost from the system becomes 
roughly a half, i.e., $\sim 30$ \% 
of the accreted mass and the mass accumulation 
efficiency is $\eta \sim 0.7$. 

Another important problem is the effect of rotation. 
\citet{yoo04} calculated He flashes including rotation and showed that  
rotation generally makes flashes milder because of a decrease in effective
gravity and contamination of C and O by rotational mixing at the base of
He layer. On the other hand, the ignition mass seems to be hardly 
affected by rotation (we judged it from their Figure 1). 
If we simply apply these results to our model, 
we may say that H flashes will be milder 
because of a decrease in effective gravity and contamination of He 
(not WD material, because heavy element enrichment is not observed), 
while the timescale of inter-flash phases would be unchanged.
Thus, the mass lost from the system could become further small 
(i.e., $\eta \gtrsim 0.7$).


\begin{figure}
\epsscale{1.1}
\plotone{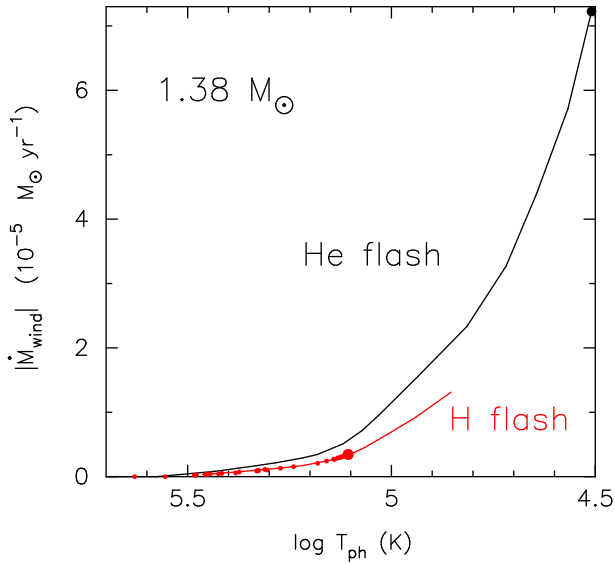}
\caption{Wind mass-loss rate $\dot M_{\rm wind}$
versus photospheric temperature $T_{\rm ph}$ for the H and He flashes. 
The solid lines denote the steady-state sequences of 
the He flash (black line) and H flash (red line). 
The filled red circles are the wind mass-loss rates 
of the H flash model for the recurrent nova M31N 2008-12a,
which are obtained consistently with the evolution model \citep{kat17sh}. 
See text for more details.
\label{dmdtT.he}}
\end{figure}

\subsection{Observational Properties of Helium Flash}
\label{sec_heflash}
Figure \ref{dmdtT.he} shows the wind mass-loss rates  
against the photospheric temperature. 
The filled red circles are 
of the evolution model of a H flash on a $1.38~M_\sun$ WD with 
$\dot M_{\rm acc}=1.6 \times 10^{-7}M_\sun$~yr$^{-1}$, 
calculated for M31N 2008-12a outbursts \citep{kat17sh}, in which 
the wind mass-loss rates are obtained from iteration process and 
are consistent with the optically thick wind acceleration. 
The large filled red circle represents the stage of the maximum 
wind mass-loss rate of the H flash, which may correspond to the stage of 
the optical/NIR maximum 
because in free-free emission, the optical/NIR magnitudes are 
in proportion to 
the square of the wind mass-loss rate \citep[Equation (9) in ][]{hac06kb}. 
The red line depicts the wind mass-loss rate 
of the steady-state sequence of optically thick wind solutions.  
The envelope mass at the large filled red circle is 
$2.2 \times 10^{-7}~M_\sun$.  

This figure also shows a steady-state sequence for a helium flash 
on a 1.38 $M_\sun$ WD with the chemical composition,
$Y=0.68$, $X_{\rm C+O}=0.2$, $X_{\rm Ne}=0.1$ and $Z=0.02$ \citep{kat04}.
The upper end of this line (large filled black circle)
represents the steady-state solution 
of the envelope mass, $7.2 \times 10^{-5}~M_\sun$, 
consistent with the ignition mass of our He flash, 
$7.5 \times 10^{-5}~M_\sun$.
The corresponding mass-loss rate is
$\sim 7.2 \times 10^{-5}~M_\sun$~yr$^{-1}$, 
about 20 times larger than $\sim 3.5 \times 10^{-6}~M_\sun$~yr$^{-1}$  
of the H flash. 
If we further assume that the proportionality constant of Equation (9) 
in \citet{hac06kb} is common among He-rich and H-rich matter, the 
free-free emission flux of He nova is 
roughly $20^2 /2/8 \sim 25$ times higher than those of the H flash, 
where the factors 2 and 8 are the difference of 
electron and nuclei number densities between the H-rich and He-rich envelopes. 
Thus, the peak magnitude in optical/NIR emission 
of He nova is $2.5 \times \log (f_{\rm He~nova}/f_{\rm H~nova}) 
=2.5 \times \log 25$=3.5 mag brighter
than that of the M31N 2008-12a outburst. 
We can expect a bright He nova outburst. 
Thus, we encourage search for a He flash in archival plates.

Figure \ref{hr} shows the sequence of the steady/static solutions 
(solid green line) that approximately represents the decay phase
of He nova if the gravitational energy release rate $L_{\rm G}$
is negligible \citep{kat17sh}.  
The SSS phase of this sequence lasts 37 days, 
from the end of the wind mass-loss (small open circle) to the left-end
of the line, where He nuclear burning ends. 
The effect of the gravitational energy release would be 
slowdown of the evolution \citep[see][]{kat17sh}. 
If this effect amounts a few ten percent,  
the SSS phase may last $\sim 40-50$ days. 
However, He burning produces substantial amount of carbon, which would 
trigger thick dust-shell formation. In the outburst of the He nova V445 Pup 
dust blackout occurs 210 days after the optical discovery, 
which prevents direct observation of the WD until now 
\citep{kam02,kat03,ash03,iij08,kat08,wou09}.
In the same way, thick dust-shell formation would possibly hinder optical
observation of a later phase of the He nova outburst of M31N 2008-12a. 

It is interesting that, in the He flash, the prompt X-ray flash lasts
much shorter (15~min) but the late SSS phase lasts longer (40 -- 50 days) 
than those in the H flash (a day and a week, respectively). 
The He flash is much stronger 
because of high temperature of He ignition, which results in a shorter 
rising time, i.e., shorter X-ray flash. 
In the decay phase, on the other hand, 
the He envelope mass is 100 times larger than that of H envelope but  
nuclear burning energy release of He is 10 times smaller. 
Thus, the late SSS phase lasts almost one order of magnitude longer 
than that in the H flash.

\section{Conclusions} 
\label{sec_conclusion}

Our main results are summarized as follows.

\noindent
1. We present 1500 consecutive hydrogen shell flashes 
on a 1.38 $M_\sun$ WD 
with a mass accretion rate of $1.6 \times 10^{-7}M_\sun$~yr$^{-1}$.  
These parameters are taken from a model of M31N 2008-12a. 
The shell flash soon reaches a steady-state only after 
$\sim 70$ cycles with a small period variation of 8\%.  
Until the He ignition, each H shell flash is almost identical 
in the photospheric luminosity and there are no noticeable 
precursors of the 
forthcoming He flash even in the epoch of the last hydrogen flash.

\noindent
2. The helium thermonuclear runaway occurs in an extremely short timescale 
compared with that of hydrogen. 
The nuclear burning rate reaches $L_{\rm nuc}=4 \times 10^{11}~L_\sun$ at 
its peak, $10^5$ times larger than that of H burning. 
Even such large energy production rates,
most of the nuclear energy is absorbed 
by the inner part of the burning region. As a result 
the photospheric luminosity $L_{\rm ph}$ is almost equal to the Eddington luminosity. 

\noindent
3. We present a prompt X-ray flash light-curve of the He nova.  
The duration of the X-ray flash of He nova is as short as 15 min, 
which makes the detection very difficult 
even in high cadence observations as done 
in the X-ray flash of the 2015 outburst 
of M31N 2008-12a \citep{kat16xflash}. 

\noindent
4. During the early phase of the He outburst, most of the surface hydrogen 
is convectively mixed into the deep interior and is burned into helium 
before the optically thick wind mass-loss occurs. 
Thus, the ejecta would contain much less hydrogen 
(i.e., $\ll 1 \times 10^{-7}~M_\sun$) than we expect from 
the amount of accreted hydrogen-rich matter before the He flash occurs.

\noindent
5. The optically thick winds begin at the end of the 
X-ray flash ($t_{\rm He}=16.1$ min), when the photospheric temperature
decreases to $\log T_{\rm ph}$ (K) $= 5.55$, owing to acceleration
by the Fe opacity peak. Characteristic properties, such as the occurrence
of the wind and interior structure of the envelope, 
are essentially the same as those in hydrogen shell flashes 
\citep{kat16xflash}.

\noindent
6. M31N 2008-12a is a promising candidate of He novae.
We expect a He nova outburst having a very short X-ray flash (15 min),
very bright optical/NIR peak ($\sim3.5$ mag brighter than M31N 2008-12a), 
much longer nova duration ($>2$ years), and longer SSS phase 
(40 -- 50 days or more). 
Thus, we encourage search for a He flash in archival plates.

\acknowledgments
M.K. and I.H. thank M. Henze for fruitful discussion on M31N 2008-12a 
and He novae. 
We thank the anonymous referee for useful comments that improved the manuscript.
 This research has been supported in part by Grants-in-Aid for
 Scientific Research (15K05026, 16K05289)
 of the Japan Society for the Promotion of Science.



















\begin{thebibliography}{}

\bibitem[Ashok \& Banerjee (2003)]{ash03}
Ashok, N. M., \& Banerjee, D. P. K. 2003, \aap,409, 1007

\bibitem[Darnley et al. (2015)]{dar15}
Darnley, M. J., Henze, M., Steele, I. A. et al. 2015,
\aap, 580, 45

\bibitem[Darnley et al. (2016)]{dar16} Darnley, M. J., Henze, M., Bode, M.F. et al. 2016, \apj, 833, 149

\bibitem[Darnley et al. (2016)]{dar16errata} Darnley, M. J., Henze, M., Steele, I. A. et al. 2015,
\aap, 593, 3 (Erratum)

\bibitem[Darnley et al. (2014)]{dar14}
Darnley, M. J., Williams, S. C., Bode, M. F., et al. 2014, \aap, 563, L9


\bibitem[Denissenkov et al. (2013)]{den13}
Denissenkov, P. A., Herwig, F., Bildsten, L., \& Paxton, B.
2013, \apj, 762, 8

\bibitem[Epelstain et al. (2007)]{epe07}
Epelstain, N.,  Yaron, O., Kovetz, A. \& Prialnik, D. 2007, \mnras, 374, 1449

\bibitem[Hachisu \& Kato (2001)]{hac01kb}
Hachisu, I., \& Kato, M. 2001, \apj, 558, 323

\bibitem[Hachisu \& Kato (2006)]{hac06kb}
Hachisu, I., \& Kato, M. 2006, \apjs, 167, 59

\bibitem[Hachisu \& Kato (2010)]{hac10k}
Hachisu, I., \& Kato, M. 2010, \apj, 709, 680

\bibitem[Hachisu \& Kato (2014)]{hac14}
Hachisu, I., \& Kato, M. 2014, \apj, 785, 97

\bibitem[Hachisu \& Kato (2015)]{hac15k}
Hachisu, I., \& Kato, M. 2015, \apj, 798, 76

\bibitem[Hachisu \& Kato (2016)]{hac16k}
Hachisu, I., \& Kato, M. 2016, \apj, 816, 26

\bibitem[Hachisu et al. (1999a)]{hkn99}
Hachisu, I., Kato, M., \& Nomoto, K. 1999a, \apj, 522, 487 

\bibitem[Hachisu et al. (1999b)]{hknu99}
Hachisu, I., Kato, M., Nomoto, K., \& Umeda, H. 1999b, \apj, 519, 314

\bibitem[Hachisu et al. (2000)]{hkkm00}
Hachisu, I., Kato, M., Kato, T., \& Matsumoto, K. 2000,
\apjl, 528, L97 


\bibitem[Hachisu et al. (2006)]{hac06b}
Hachisu, I., Kato, M., Kiyota, S., et al. 2006, \apjl, 651, L141

\bibitem[Hachisu et al. (2016)]{hac16sk}
Hachisu, I., Saio, H., \& Kato, M. 2016, \apj, 824, 22

\bibitem[Henze et al. (2014)]{hen14}
Henze, M., Ness, J.-U., Darnley, M., et al. 2014, \aap, 563, L8

\bibitem[Henze et al. (2015a)]{hen15}
Henze, M., Ness, J.-U., Darnley, M., et al. 2015, \aap, 580, 46

\bibitem[Henze et al. (2015b)]{hen15.half.period} Henze, M., Darnley, M. J.,
Kabashima, F., et al. 2015b, \aap, 582, 8

\bibitem[Hillman et al. (2016)]{hil16} Hillman, Y., Prialnik, D., Kovetz, A.\& Shara, M. M. 2016, \apj, 819, 168 

\bibitem[Iben (1982)]{ibe82}
Iben, I., Jr. 1982,  \apj, 259, 244

\bibitem[Iben et al. (1983)]{ibe83} Iben, I. Jr. , Kaler, J. B., Truran, J. W.\& Renzini, A. 1983, \apj, 264, 605
  

\bibitem[Idan et al. (2013)]{ida13} Idan, I., Shaviv, N. J., \& Shaviv, G. 2013, \mnras, 433, 2884

\bibitem[Iijima \& Nakanishi (2007)]{iij08}
Iijima,T., \& Nakanishi, H. 2008, \aap, 482, 865


\bibitem[Kamath \& Anupama (2002)]{kam02}
Kamath, U. S., \& Anupama, G. C. 2002, Bull. Astr. Soc. India, 30, 679

\bibitem[Kato (1966)]{kat66} Kato, S. 1966, \pasj, 18, 374

\bibitem[Kato \& Kanatsu (2000)]{kan00} Kato, T., \& Kanatsu, K., 2000, IAU Circ. 7552

\bibitem[Kato \& Hachisu (1994)]{kat94h}
 Kato, M., \& Hachisu, I., 1994, \apj, 437, 802

\bibitem[Kato \& Hachisu (2003)]{kat03} Kato, M., \& Hachisu, I. 2003, \apj, 598, L107


\bibitem[Kato \& Hachisu (2004)]{kat04} Kato, M., \& Hachisu I. 2004, \apjl, 613, 129

\bibitem[Kato et al.(2008)]{kat08}
Kato, M., Hachisu, I., Kiyota, S., Saio, H., 2008, ApJ, 684, 1366

\bibitem[Kato et al. (1989)]{kat89} Kato, M., Saio, H., \& Hachisu, I. 1989, \apj, 340, 509

\bibitem[Kato et al. (2015)]{kat15sh}
Kato, M., Saio, H., \& Hachisu, I. 2015, \apj, 808, 52,

\bibitem[Kato et al. (2017)]{kat17sh}
Kato, M., Saio, H., \& Hachisu, I. 2017, \apj, 838, 153

\bibitem[Kato et al. (2014)]{kat14shn}
Kato, M., Saio, H., Hachisu, I., \& Nomoto, K. 2014, \apj, 793, 136


\bibitem[Kato et al. (2016)]{kat16xflash} Kato, M., Saio, H., Henze, M. et al. 2016,
\apj, 830, 40


\bibitem[Kovetz (1998)]{kov98}
Kovetz, A. 1998, \apj, 495, 401

\bibitem[Nariai et al. (1980)]{nar80}
 Nariai, K., Nomoto, K., \& Sugimoto, D. 1980, \pasj, 32, 473

\bibitem[Nomoto et al. (1979)]{nom79} Nomoto, K., Nariai, K., \& Sugimoto, D. 1979,
\pasj, 31, 287


\bibitem[Prialnik (1986)]{pri86}
Prialnik, D. 1986, \apj, 310, 222

\bibitem[Prialnik \& Kovetz (1995)]{pri95}
Prialnik, D., \& Kovetz, A. 1995, \apj, 445, 789

\bibitem[Regev \& Shara (1989)]{reg89} Regev, O. \&  Shara, M. 1989, \apj, 340,1006


\bibitem[Shakura \& Sunyaev (1973)]{sha73} Shakura, N. I., \& Sunyaev, R. A. 1973, \aap, 24, 337


\bibitem[Sion et al. (1979)]{sio79}
Sion, E. M., Acierno, M. J., \& Tomczyk, S. 1979, \apj, 230, 832

\bibitem[Sparks et al. (1978)]{spa78} Sparks, W.N., Starrfield, S., \& Truran, J.W.
1978, \apj, 220, 1063

\bibitem[Spruit (1992)]{spr92} Spruit, H.C. 1992, \aap, 253, 131

\bibitem[Tang et al. (2014)]{tan14}
Tang, S., Bildsten, L., Wolf, W. M., et al. 2014, \apj, 786, 61

\bibitem[Wolf et al. (2013a)]{wol13a}
Wolf, W. M., Bildsten, L., Brooks, J., \& Paxton, B. 2013a, \apj, 777, 136

\bibitem[Wolf et al. (2013b)]{wol13b}
Wolf, W. M., Bildsten, L., Brooks, J., \& Paxton, B. 2013b, \apj, 782, 117,
(Erratum)



\bibitem[Woudt et al. (2009)]{wou09}
Woudt, P.A., et al., 2009, ApJ, 706, 738

\bibitem[Yaron et al. (2005)]{yar05} 	
Yaron, O., Prialnik, D., Shara, M.M., \& Kovetz, A. 2005, \apj, 623, 398

\bibitem[Yoon et al. (2004)]{yoo04} Yoon, S.-C., Langer, N., \& Scheithauer, S. 2004, \aap, 425, 217

\end{thebibliography}
\end{document}